\documentstyle[12pt]{article}
\newlength{\extraspace}
\newlength{\extraspaces}
\newcommand{\newsection}[1]{
\vspace{15mm}
\pagebreak[3]
\addtocounter{section}{1}
\setcounter{subsection}{0}
\setcounter{footnote}{0}
\begin{flushleft}
{\large\bf \thesection. #1}
\end{flushleft}
\nopagebreak
\medskip
\nopagebreak}

\newcommand{\newsubsection}[1]{
\vspace{1cm}
\pagebreak[3]

\addtocounter{subsection}{1}
\noindent{ \large \bf #1}
\nopagebreak
\vspace{2mm}
\nopagebreak}

\newcommand{\F}{F_{h_- , h_+}}
\newcommand{\dy}{\frac{d \Omega}{dy}}

\begin{document}
\addtolength{\baselineskip}{.7mm}

\input epsf

\thispagestyle{empty}

\begin{flushright}
{\sc PUPT}-1493\\
April 1995
\end{flushright}
\vspace{.3cm}

\begin{center}
{\Large\bf{Static and Dynamic Analysis of a Massless Scalar
 \\[2mm]   Field Coupled with a Class of Gravity Theories
}}\\[10mm]
{\sc Youngjai Kiem}\\[3mm]
{\it Joseph Henry Laboratories\\[2mm]
 Princeton University\\[2mm]
Princeton, NJ 08544\\[2mm] }
{\sc And}\\[3mm]
{\sc Dahl Park}\\[3mm]
{\it Department of Physics\\[2mm]
 KAIST\\[2mm]
Taejon 305-701, KOREA\\[2mm]
E-mail: dpark@chep5.kaist.ac.kr \\
        ykiem@puhep1.princeton.edu}
\\[10mm]

{\sc Abstract}
\end{center}

General static solutions for a massless scalar field coupled to
a class of effectively 2-d gravity theories continuously
connecting spherically symmetric $d$-dimensional Einstein gravity
($d >3$) and the CGHS model are analytically obtained.  They include
black holes and point scalar charge solutions with naked
singularities, and are used to give an analytic proof of no-hair
theorem.  Exact scattering solutions in $s$-wave 4-d Einstein gravity
are constructed as a generalization of corresponding static
solutions.  They show the existence of black hole formation threshold
for square pulse type incoming stress-energy flux, above which trapped
surfaces are dynamically formed.  The relationship between this
behavior and the numerically studied phase transition in this system
\cite{choptuik} is discussed.

\noindent

\vfill

\newpage

\newsection{Introduction}
     Recently, there has been remarkable progress in black
hole physics, as we gain valuable insights from model
gravity theories such as the CGHS model where the theory is
exactly integrable \cite{CGHS}.  In these simplified settings, we can
address such vexing questions as the presumed information
loss via Hawking radiation process, black hole thermodynamics
and related issues in a much more concrete analytical way.
Even some developments on quantization were manageable \cite{giddings}.
At the same time, progress has also been made in 4-d Einstein
gravity coupled with various matter fields.  Especially interesting
approach, to name an example, is the study of the gravitational
collapse in spherically
symmetric reduction of 4-d Einstein gravity coupled with a single
massless scalar field \cite{chris}.  Several rigorous mathematical
results and some explicit solutions were obtained so far \cite{dc}
\cite{roberts}.  Moreover, an intriguing critical behavior
was observed in this system near the onset of the black hole
formation through numerical analysis \cite{choptuik}.
A generic 1-parameter class of solutions $S[p]$ decomposes into two
phases, depending on the magnitude of $p$ that measures the strength of
self-gravitational interaction.  For $p < p^*$, where $p^*$ is the
critical value, the gravitational
collapse of incoming stress-energy flux is followed by an explosion,
reflecting back the flux toward the future infinity.  If $p > p^*$, the
black holes are dynamically formed and the mass of
the asymptotically static
black hole $M_{BH}$ shows a universal scaling behavior,
$M_{BH} \simeq | p - p^* |^{\Delta}$ with $\Delta \simeq 0.37$.
The analytic explanation for this phase transition has been only partially
successful \cite{brady} \cite{kiem}, although in the context of
the (1-loop corrected) CGHS model the analytic explanation for the
similar phase transition with $\Delta  = 0.5$ is available
\cite{strominger}.

     As has been well established now, the spherically symmetric
4-d Einstein gravity and the CGHS model, both of them
(effectively) 2-dimensional,
can be given a unified treatment as a generalized 2-d dilaton
gravity theory.  To be specific, we can
construct a class of 2-d gravity theories that continuously connect
these two interesting theories \cite{birkhoff}.
In view of this point, our interest
in analyzing these interconnecting theories is at least two-folds;
first, we would like to know to exactly what extent many interesting
results obtained for the CGHS model are the generic lessons on
gravitational physics and what results are artifacts of the particular
choice of parameters.  Secondly, since
the explicit analytic calculations in 4-d Einstein gravity are very
difficult, we want to find a simpler setting to answer many
classical (and possibly quantum) questions in general relativity
in detail. This can possibly teach us more about classical
and quantum cosmological questions such as the realistic gravitational
collapsing process in general relativity.

     Keeping these considerations in mind, in this paper,
we systematically study classical general static solutions
for a scalar field coupled with various effectively 2-d gravity
theories that continuously connect spherically symmetric 4-d
Einstein gravity and the CGHS model.
The model action we consider
here is given as follows
\begin{equation}
I = \int d^2 x \sqrt{- g} e^{ -2 \phi} ( R^{(2)} +  \gamma
g^{\alpha \beta} \partial_{\alpha} \phi
\partial_{\beta} \phi
+ \mu e^{2 \lambda \phi} - \frac{1}{2}
e^{-2 \phi (\delta -1 ) } g^{\alpha \beta}
\partial_{\alpha} f \partial_{\beta} f  )
\label{action}
\end{equation}
and we focus on the case of $\delta = 1$ for analytic tractability.
Here $R^{(2)}$ denotes the 2-d scalar curvature.  $\phi$ and
$f$ represent a dilaton field and a massless scalar field,
respectively.  $\gamma$, $\lambda$, $\delta$ and
$\mu$ are real parameters.  The specific choice of these
4 parameters corresponds to a particular gravity theory.
Originally, the action introduced above was considered as a
2-d target space effective action resulting from string
theory, if we neglect loop corrections.
Other interpretations
of the action were found since then.
After $(d-2)$-dimensional angular integration, for example,
the action for
spherically symmetric $d$-dimensional Einstein gravity theories
reduces to Eq.(\ref{action}), modulo total derivative terms, with $\gamma
= 4 (d-3)/(d-2) $, $\lambda = 2/(d-2)$ and $\delta = 1$.  The
dilaton field $\phi$ is related to the geometric radius $r$
of the transversal $(d-2)$-sphere as $r = \exp ( - 2 \phi / (d-2) )$.
On the other hand, the CGHS model
is recovered if we set $\gamma = 4$,
$\lambda = 0$ and $\delta = 0$.   Thus, we see that as long as
the pure gravity sector is concerned, the CGHS model is exactly the
same as $d \rightarrow \infty$ limit of $d$-dimensional Einstein
gravity.  The only difference is the dilaton prefactor in
the action for the scalar field in the latter theory.

     In section 2, we derive the general static solutions
for our model with $\delta = 1$ using a
rather elementary method.  Specifically, for general value
of $\delta$, we integrate the equations of motion, a
system of highly non-linear second-order coupled differential
equations, by finding the relevant number of symmetries of the
action, reducing the order of differential equations by one.
Thus, the result of
integration can be represented as the conservation
of several Noether charges.
These equations are more analytically tractable
and, for $\delta = 1$ and $2- \lambda - \gamma/4 > 0$, i.e. $ d>3$,
they can be further integrated
exactly to yield closed form expressions.  The main novelty here
is our method of derivation that enables us to get the {\it general}
static solutions.  Unlike a well-developed technique of generating
solutions for complicated gravity theories from vacuum solutions
of Einstein equations, our method solves equations of motion
directly, thereby getting general solutions in a straightforward
and simple fashion.  Our analysis shows that the space
of all possible static solutions modulo coordinate
transformation is two
dimensional (or several sheets of two dimensional space).
The space is parameterized, roughly, by black hole mass and
scalar charge.  For non-vanishing black hole mass, in 4-d
Einstein gravity, we reproduce the exact solutions obtained by
Janis {\it et.al.}\cite{janis} long time ago using a solution generating
technique from the vacuum solutions of Einstein equations.
Recalling that solutions we get here are general
static solutions, our further analysis leads to a new analytic
proof of no-hair theorem that takes the backreaction on space-time
geometry non-perturbatively into account.  More interesting
result is the case of vanishing black hole mass.  In this case,
there is a general relativistic analog of a (non-relativistic)
point scalar charge
solution in the vacuum space-time in each theory.  Particularly, these
solutions contain naked singularities.
Although they are solutions for which the total
gravity-matter action in $d$-dimensional Einstein gravity
\[ I^{(d)} = \int d^d x \sqrt{- g^{(d)} } (R^{(d)} - \frac{1}{2}
g^{(d) \alpha \beta} \partial_{\alpha} f \partial_{\beta} f ) \]
vanishes, it remains to be seen whether
they are stable against classical and quantum perturbations.

     In section 3, we find the extension of the static
point scalar charge solutions into scattering solutions is easy
in some cases, namely, the spherically symmetric reduction
of 4-d Einstein gravity.  Utilizing the extension, we construct
dynamic solutions, which demonstrate that, for the square pulse
type incoming stress-energy flux, there exists a black hole
formation threshold for the incoming stress-energy flux $T_{in}$.
Specifically, in our case, the threshold value
of the incoming stress-energy flux, $T_{in}^c$, that
plays the role of $p^*$ is found to be
\[ \lim_{r \rightarrow \infty} \int_{angle} T_{in}^c /c^2
    = 4 \pi T_{in}^c / c^2
    = \frac{c^3}{4G} , \]
where $G$ is the gravitational constant
and $c$ is the speed of light.
If $T_{in} > T_{in}^c$,
black holes are dynamically formed (supercritical
case, in the language of phase transition).  Then, an injection of shock
wave type stress-energy flux naturally belongs to the supercritical
case, for the value of $T_{in}$ is very large during a
very short period of time, and the corresponding solutions are
constructed.  Below the threshold, $T_{in} < T_{in}^c$,
the incoming pulse type stress-energy flux gets reflected forward into the
future null infinity (subcritical case).
In supercritical case, we compute the critical
exponent for our solutions taking the apparent mass of the
resulting black hole immediately after the formation as an
order parameter, getting $M_A \simeq |p - p^*|^{0.5}$.  We further indicate
that our analysis here supports the aforementioned numerical
study \cite{choptuik} by showing a plausible mechanism that generates
a non-trivial scaling relation between $M_A$ and $M_{BH}$.
The difficulties of the similar extension in other generic gravity
theories are then discussed.  Based on the qualitative arguments,
we later conjecture that the scaling relation for $d$-dimensional
Einstein gravity for $d>4$ is $M_{BH} \simeq |p - p^* |^{0.5}$, unlike
the 4-dimensional (and 3-dimensional \cite{evans}) results.  This
dependence of the critical exponent on the space-time dimensionality
is very similar to the well-known cases in condensed matter physics,
such as the ferromagnetic transition.

     In section 4, we discuss our results and their possible physical
implications in various physical contexts.

\newsection{General Static Solutions}

     We start by deriving general static solutions in conformal
gauge.  The solutions we get in this section can also be obtained
under a different gauge choice as sketched in Appendix.  The calculations
in Appendix show that, in case of spherically symmetric 4-d
Einstein gravity coupled with a massless scalar field, the solutions
for non-vanishing black hole mass are identical to those found
in \cite{janis} using a solution generating technique.

\newsubsection{2.1. Derivation of General Static Solutions}

     The equations of motion we have to solve to get the
static solutions are obtained from our action by varying it
with respect to the metric tensor, dilaton field and
the massless scalar field;
\begin{equation}
D_{\alpha} D_{\beta} \Omega - g_{\alpha \beta} D \cdot
D \Omega + \frac{\gamma}{8} ( g_{\alpha \beta}
\frac{ (D \Omega )^2 }{\Omega} - 2
\frac{ D_{\alpha} \Omega D_{\beta} \Omega }{\Omega} )
+ \frac{\mu}{2} g_{\alpha \beta} \Omega^{1 - \lambda}
\label{geom}
\end{equation}
\[ + \frac{1}{2} \Omega^{\delta} D_{\alpha} f
   D_{\beta} f  - \frac{1}{4} \Omega^{\delta}
g_{\alpha \beta} (Df)^2 = 0 \]
\begin{equation}
R + \frac{\gamma}{4} ( \frac{ (D \Omega )^2}{\Omega^2}
-2 \frac{D \cdot D \Omega}{\Omega } ) + ( 1- \lambda )
\mu \Omega^{- \lambda} - \frac{\delta}{2} \Omega^{\delta -1}
  (D f)^2 = 0
\end{equation}
\begin{equation}
 g^{\alpha \beta} D_{\alpha} ( \Omega^{\delta}
D_{\beta} f ) = 0 ,
\end{equation}
where we define $\Omega = e^{-2 \phi}$ and $D$ denotes
the covariant derivative.  Since the only matter coupling
in our case is a massless scalar field,
we choose to work in a conformal
gauge.  Thus, we write the metric $g_{\alpha
\beta} = - \exp ( 2 \rho + \frac{\gamma}{2} \phi )
dx^+ dx^- $.  Our convention, for the calculational
simplicity, is the negative signature for space-like coordinates
and the positive signature for a time-like coordinate.
In spherically symmetric 4-d Einstein gravity, for example, this
implies $\mu = -2$ using Gauss-Bonnet theorem.
Under this choice of coordinates, the original action
gets simplified to become
\begin{equation}
I = \int dx^+ dx^- ( 4 \Omega \partial_+ \partial_-
\rho + \frac{\mu}{2} e^{2 \rho } \Omega^{1 - \lambda -
\gamma / 4} + \Omega^{\delta} \partial_+ f
\partial_- f    ),
\label{caction}
\end{equation}
modulo total derivative terms.  In our choice of the conformal
factor, we included a contribution from the dilaton field.
This contribution, up to a total derivative term we
threw away, was so chosen to
cancel the kinetic energy term for the dilaton
field, rendering a simplified form of the action.
Eq.(\ref{caction}) should be supplemented by gauge constraints
resulting from the choice of the conformal gauge.  They
are calculated to be
\begin{equation}
\partial_{\pm}^2 \Omega - 2 \partial_{\pm} \rho
\partial_{\pm} \Omega + \frac{1}{2}
\Omega^{\delta} ( \partial_{\pm} f )^2 = 0 ,
\end{equation}
  from Eq.(\ref{geom}) for $g_{\pm \pm}$ components of
the metric tensor.

     The general static solutions for the equations of motion
from the action (\ref{caction}) under the gauge constraints
can be found as follows; we can consistently reduce
the partial differential equations (PDE's)
into the coupled second
order ordinary differential equations (ODE's) by
assuming all functions depends on a single space-like coordinate
$x = x^+ x^-$ \footnote{This coordinate becomes space-like due the
signature choice made here}.  Although it is not, in general,
possible to have a
consistent reduction of PDE's to ODE's assuming an arbitrary
one-variable dependence, our choice turns out to be
consistent.  This procedure yields the following ODE's
\begin{equation}
x \ddot{\Omega} + \dot{\Omega}
+ \frac{\mu}{4} e^{2 \rho} \Omega^{1 - \lambda - \gamma /4}
\dot{\rho} = 0 ,
\label{seom}
\end{equation}
\[
x \ddot{\rho} + \dot{\rho} + \frac{\mu}{8}
( 1 - \lambda - \gamma /4 ) \frac{e^{2 \rho}}
{ \Omega^{\lambda + \gamma /4 }}  +
\frac{\delta}{4} x \Omega^{\delta -1} \dot{f}^2
= 0 , \]
\[
  \frac{d}{dx} ( x \Omega^{\delta} \dot{f} ) = 0 , \]
along with the gauge constraint
\[ \ddot{\Omega} - 2 \dot{\rho} \dot{\Omega} +
\frac{1}{2} \Omega^{\delta} \dot{f}^2 = 0 , \]
where the dot represents taking a derivative with respect
to $x$.  The complete general solutions of the above
ODE's are the same as the general static Gravity-Scalar
solutions under a particular choice of the conformal
coordinates.  The solutions under a different choice of
conformal coordinates can be obtained from the solutions
of Eqs.(\ref{seom}) by proper conformal transformations
of $x^{\pm}$.  If we set $f = 0$, i.e., if we consider the
pure gravity sector, Birkhoff's theorem ensures that the
solutions obtained in this fashion are indeed general solutions
of the original PDE's.  The equations of motion other than
the gauge constraint can be summarized by an
action
\begin{equation}
 I = \int dx ( x \dot{\Omega} \dot{\rho}
- \frac{\mu}{8} e^{2 \rho} \Omega^{1 - \lambda - \gamma /4}
- \frac{1}{4} x \Omega^{\delta} \dot{f}^2 ) .
\label{eaction}
\end{equation}
One can straightforwardly verify that by varying this
action with respect to $\Omega$, $\rho$ and $f$, we
recover Eqs.(\ref{seom}).

     We observe that the action (\ref{eaction}) has two
obvious rigid continuous symmetries.  First symmetry is clear,
for $f$ field appears only through its first derivative.  Thus,
we see that $f \rightarrow f + \alpha$ is a symmetry for
an arbitrary constant $\alpha$.  The second symmetry is
the transformation $ x \rightarrow x e^{\alpha}$
and $\rho \rightarrow \rho - \alpha /2$.  This symmetry
represents the scaling invariance of the action (\ref{eaction}).
The existence of this symmetry is necessary since the original
action (\ref{caction}) is invariant under the (local) conformal
transformations.  Since there are three functions
$\rho$, $\Omega$ and $f$ that we should solve in terms of
$x$, we can integrate Eqs.(\ref{seom}) once to reduce them
to first order ODE's if we can find one additional rigid
continuous symmetry of the action.
The remaining symmetry turns out to be
$ x \rightarrow x^{1 + \alpha} $, $\rho \rightarrow
\rho - (2 - \lambda - \gamma /4 ) \ln ( 1 + \alpha ) /2
- \alpha \ln x /2$, $\Omega \rightarrow \Omega (1 + \alpha )$,
and $f \rightarrow f (1 + \alpha )^{(1-\delta ) /2}$.
This transformation changes
the action (\ref{eaction}) by a total derivative. We can deduce
the form of this symmetry by the following physical consideration;
the asymptotically flat spatial coordinate at spatial infinity
is related to the spatial coordinate that is flat near the
black hole horizon by the conformal transformation $x^{\pm}
\rightarrow \ln x^{\pm}$,
as is familiar from the definition of the tortoise
coordinate.  The physical system in our consideration is also scale
invariant in this asymptotically flat region.  Our third symmetry
is this logarithmic conformal coordinate transformation followed
by rescaling of $\Omega$ and $f$ fields, representing this
asymptotic scale invariance.

     Given these three symmetries, we can construct the
corresponding Noether charges.
\begin{equation}
   f_0 = x \Omega^{\delta} \dot{f}
\label{f0}
\end{equation}
\begin{equation}
   c_0 = x^2 \dot{\rho} \dot{\Omega} + \frac{1}{2} x
\dot{\Omega} - \frac{1}{4} x^2 \Omega^{\delta} \dot{f}^2
+ \frac{\mu}{8} x e^{2 \rho} \Omega^{1 - \lambda
- \gamma /4 }
\label{c0}
\end{equation}
\begin{equation}
   M  + 2 c_0 \ln x = 2 x \dot{\rho} \Omega -
(2 - \lambda - \gamma /4) x \dot{\Omega}
+  \Omega + \frac{\delta -1}{2} x \Omega^{\delta}
f \dot{f}
\label{c1}
\end{equation}
We can rewrite Eqs.(\ref{seom}) in a form that represents
the conservation of Noether charges $f_0$, $c_0$ and $M$.
By integrating this form of equations of motion, we get
the equations shown above where $f_0$, $c_0$ and $M$ are
constants of integration.  Additionally, the gauge constraint
reduces to a condition $c_0 = 0$.  In the absence of matter
fields, modulo coordinate transformations, the general
solutions of gravity theories considered here are parameterized
by a single parameter, i.e., the black hole mass.
Similarly, in our case the gauge
constraint kills a redundant degree of freedom, $c_0$, for
gravity sector.  As will be clear in the following sections,
two remaining Noether charges $f_0$ and $M$ can be related
to a scalar charge and a black hole mass, respectively.

     We can further solve the above equations and this
process can be straightforwardly carried out for $\delta = 1$.
In this case, we can solve $\rho$ from Eq.(\ref{c1})
and $f$ from Eq.(\ref{f0}) to find
\begin{equation}
\rho = \frac{M}{2} \int \frac{dy}{\Omega} + \frac{1}{2} (1+ q)
\ln \Omega - \frac{1}{2} y +  \rho_0
\label{rho}
\end{equation}
and
\begin{equation}
f = f_0 \int \frac{dy}{\Omega} + f_1  ,
\label{f}
\end{equation}
where we define $y = \ln x $ and $q = 1- \lambda - \gamma /4$.
Here $\rho_0$ and $f_1$ are additional constants of integration.
For spherically symmetric reduction of $d$-dimensional Einstein
gravity, we have $1 + q = (d-3)/(d-2)$.
Plugging Eqs.(\ref{rho}) and (\ref{f}) into Eq.(\ref{c0}), we
get a decoupled equation for $\Omega$, an integro-differential
equation.
\begin{equation}
M  \frac{d\Omega}{dy} +  (1 + q) ( \frac{d\Omega}{dy}
)^2 - \frac{1}{2} f_0^2 = - \frac{\mu}{4} \Omega^{2 + 2q}
\exp  ( M  \int \frac{dy}{\Omega} + 2 \rho_0 )
\label{intdif}
\end{equation}
Our goal is to solve this equation to get as explicitly as possible
the closed form expression of $\Omega$ in terms of $y$.
By differentiating it with respect to $y$
we have
\begin{equation}
\frac{d^2}{dy^2} \Omega = ( M  \frac{d}{dy} \Omega
+ (1 + q) ( \frac{d}{dy} \Omega )^2 - \frac{1}{2}
f_0^2 ) \frac{1}{\Omega} ,
\label{interim}
\end{equation}
a second order ODE.  Integrating Eq.(\ref{interim}) once
is immediate;
we find
\begin{equation}
k \Omega^{1+q} =
    |  h_+ -  \frac{d\Omega}{dy} |^{\frac{h_+}{h_+ + h_- }}
    |  h_- +  \frac{d\Omega}{dy} |^{\frac{h_-}{h_+ + h_- }}
\label{omega1}
\end{equation}
\[ \equiv  F_{h_- , h_+} ( \frac{d \Omega}{dy} ) \]
where we define a function $F_{h_- , h_+}$ and two
numbers
\[ h_{\pm} = \frac{1}{2(1+q)} ( \sqrt{M^2 + 2 (1 +q ) f_0^2}
\mp M )  \ge 0 \]
are introduced.
We note that $k$ is a constant of integration introduced when we go
from Eq.(\ref{interim}) to Eq.(\ref{omega1}).  Under the assumption
$1+q > 0$, the asymptotic analysis of Eq.(\ref{omega1}) shows that
the resulting space-time is asymptotically flat. In this situation,
a rather complicated intro-differential equation (\ref{intdif})
can be replaced by much simpler Eq.(\ref{omega1}).  Further asymptotic
analysis shows that the equivalence is insured if we identify
\[
k =  ( - \frac{\mu e^{2 \rho_0}}{4 (1 + q)} )^{1/2}
\]
If we properly choose the range of $\Omega$ and
$\frac{d \Omega}{dy}$, we can define the inverse of the function
$F_{h_- , h_+}$.  Then, the integration of Eq.(\ref{omega1}) is
trivially performed to yield
\begin{equation}
 \ln (x / x_0 ) = \int d \Omega \frac{1}
{ F_{h_- , h_+ }^{-1} ( k \Omega^{1 + q } )  }
\label{omega}
\end{equation}
where $x_0$ is a constant of integration.

     Eqs.(\ref{rho}), (\ref{f}) and (\ref{omega}) are general static
solutions of our problem.  Under our gauge choice, our solutions
are parameterized by 5 parameters, $M$, $f_0$, $\rho_0$,
$f_1$ and $x_0$, since the gauge constraint mandates $c_0 = 0$.
Among these, $f_1$ is just an addition of a constant term to $f$,
which is trivial.  The constants of integration $\rho_0$ and
$x_0$ represent the degree of freedom in the choice of coordinate
system, namely, the global scale choice and the reference time choice,
respectively.  Thus, modulo coordinate transformations and
the trivial $f_1$ part, we find the
general static solutions of Gravity-Scalar action in our consideration
are parameterized by two parameters $f_0$, the scalar charge, and
$M$, which will be shown to be related to the black hole mass.

\newsubsection{2.2. No-hair Theorem and Point Scalar Charges}

     In our further consideration of general static solutions,
we focus on the cases when $M \ge 0$ and $f_0 \ge 0$.  The
restriction on $M$ is motivated by the fact that we want the
black hole mass to be positive semi-definite.\footnote{In fact,
all the solution with $M<0$ contains unphysical naked
singularities as one can convince oneself from the general static
solutions.}   The results
for negative values of $f_0$ can be trivially obtained, as is clear
from Eqs.(\ref{rho}), (\ref{f}) and (\ref{omega1}).

     The derivation in the previous section shows that getting
the inverse function of $\F$ is necessary to determine the relationship
between $\Omega$ and $y = \ln x$ (note that we have $x = x^+ x^-$
in our choice of conformal coordinates).  Under our restrictions,
there are four distinctive cases we have to consider as shown in
Figs.1a - 1d for the shape of the function $\F$.  In each figure,
the physical region to take the inverse of $\F$ is shown; since the
value of $\F$ relates to some power of geometric radius of
transversal sphere, via Eq.(\ref{omega1}), we require it to vary from
zero to infinity.  Furthermore, we require $\dy$ to be positive
for large values of $\Omega$, thereby, $\F$.
In other words, $\F = 0$ is a natural space-time boundary
because it represents a space-time point where the transversal
sphere collapses to a point.  Similarly the value
$\F = \infty$ corresponds to asymptotic spatial infinities.
In every figure, $\F$ grows linearly for large, positive $\dy$,
showing the spatial infinity regions are asymptotically flat.
Thus, our choice of physical region is tantamount to considering
only static solutions with asymptotically flat space-time region
at spatial infinities and with the conventional choice for the
orientation of mappings between $\Omega$ and conformal coordinates
there.  By making this choice of physical region, we are no longer
considering parts of general static solutions that have compact
range of geometric radius.  The existence of them is clear from
Figs.1c and 1d.  These solutions with no asymptotic infinities
can be potentially important in some physical
settings such as cosmological considerations.  In this note, however,
since we will eventually be interested in constructing scattering
solutions with asymptotic in/out-regions, we restrict our attention
to space-time geometries with flat asymptotic infinities.

\begin{center}
\leavevmode
\epsfysize=5cm
\epsfbox{fig1a.eps}
\end{center}
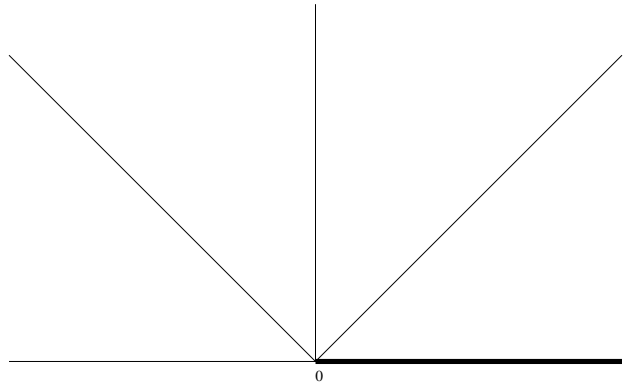
{\small Fig.1a.  The plot of $\F$ as a function of $\dy$ for
$h_+ = h_- = 0$.  The region to take the inverse of $\F$ is depicted
by a bold line.}

\begin{center}
\leavevmode
\epsfysize=5cm
\epsfbox{fig1b.eps}
\end{center}
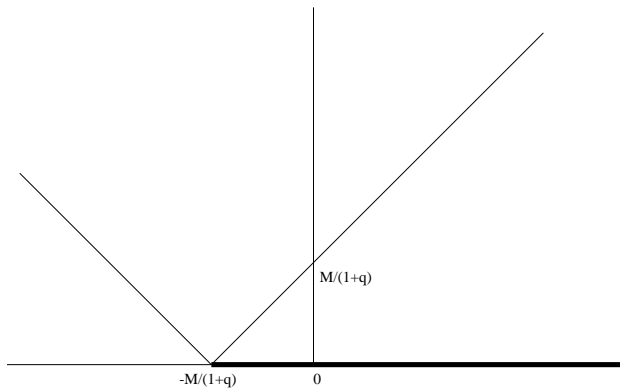
{\small Fig.1b.  The plot of $\F$ as a function of $\dy$ for
$h_+ = 0$ and $h_- = M/(1+q)$, a black hole geometry with mass $M$.
The region to take the inverse of $\F$, a bold line, contains
the point $\dy =0$. }

     In Fig.1a, we have $M=0$ and $f_0 = 0$, implying $h_{\pm} = 0$.
Then the scalar field $f$ vanishes identically and this is the
case that corresponds to the vacuum solution in each theory.
For non-vanishing value of $q$, we have
\begin{equation}
\Omega = ( - qk (y - y_0 ) )^{-1/q} .
\label{qmin}
\end{equation}
For $d$-dimensional Einstein gravity, where we have $q = - 1 /(d-2)$,
this solution represents the expression for the geometric radius
in term of conformal coordinates.  For example, in 4-d case, the
above expression becomes $r^2 = ( (\ln x^+ + \ln x^- )/2 )^2$
while the examination of Eq.(\ref{rho}) shows that $\ln x^{\pm}$
are asymptotically flat conformal coordinates.  For $q=0$, we find
\begin{equation}
\ln \Omega = k ( y- y_0 ) ,
\end{equation}
a familiar linear dilaton vacuum in pure gravity sector of the CGHS model,
as can be easily seen after a conformal transform $(x^{\pm} )^k
\rightarrow x^{\pm}$.  We can either get this expression from
Eq.(\ref{omega1}) or from Eq.(\ref{qmin}) taking $q \rightarrow
0$ limit after replacing $y_0$ with $ y_0 + 1/(qk)$.

     The black hole solutions are recovered if $M > 0$ and
$f_0 = 0$, the case in Fig.1b.  In this case, we have $h_+ = 0$
and $h_- = M/(1+q)$ and the scalar field $f$ vanishes identically.
The existence of the apparent horizon,
which is the space-time point that satisfies $\partial_+ \Omega = 0$
and, furthermore, is the same as the global event horizon in static
analysis, is clear from Fig.1b.  The indicated physical region to define
the inverse of $\F$ includes a point $\dy = 0$.  The relation
between $\Omega$ and $y = \ln x$ is calculated to be
\begin{equation}
\int \frac{d \Omega}{k \Omega^{1 + q} - M/(1+q) } = \int dy
\end{equation}
from Eq.(\ref{omega}), which reduces to
\[ \frac{2}{k} ( r + \frac{2M}{k} \ln | \frac{k}{2M} r -1  | )
= y - y_0 \]
in 4-d Einstein gravity (where $q = - 1/2$ and $\Omega = r^2$) and
\[ \Omega = e^{k (y -y_0 ) } + \frac{M}{k} \]
in the CGHS model (where $q = 0$), reproducing well-known
results.  (This result should
be carefully compared with the literature due to our somewhat
unconventional signature choice.)

\begin{center}
\leavevmode
\epsfysize=5cm
\epsfbox{fig1c.eps}
\end{center}
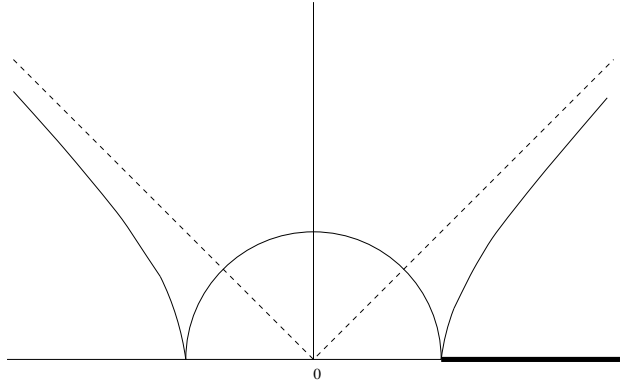
{\small Fig.1c.  The plot of $\F$ as a function of $\dy$ for
$h_{\pm} = f_0 / \sqrt{2(1+q)}$.  Dotted lines denote a
vacuum space-time.  The bold line depicts the region
to take the inverse of $\F$.}

\begin{center}
\leavevmode
\epsfysize=5cm
\epsfbox{fig1d.eps}
\end{center}
{\small Fig.1d.  The plot of $\F$ as a function of $\dy$ for
a typical $h_- > h_+ > 0$ case.  Dotted lines denote a
black hole geometry with mass $M$.  The bold line
shows the region to take the inverse of $\F$.}

     In fig.1c, we have $M=0$ and $f_0 > 0$, resulting
$h_{\pm} = f_0 / \sqrt{2(1+q)}$.  The solutions in this case
are given by
\begin{equation}
\int \frac{d \Omega}{\sqrt{f_0^2 / (2 + 2q) + k^2 \Omega^{2+2q} }}
= \int dy
\label{psc}
\end{equation}
for $\Omega$ and
\begin{equation}
f = \frac{1}{ \sqrt{2 (1+q) }} \ln (
\frac{\sqrt{2 (1+q) k^2  \Omega^{2+2q} + f_0^2 } - f_0 }
     {\sqrt{2 (1+q) k^2  \Omega^{2+2q} + f_0^2 } + f_0 } ) + f_1
\label{pscf}
\end{equation}
\begin{equation}
\rho = \frac{1}{2} (1+q) \ln \Omega - \frac{1}{2} y + \rho_0
\label{pscr}
\end{equation}
for $f$ and $\rho$, respectively.  The integral in Eq.(\ref{psc}) can
straightforwardly be represented in terms of incomplete beta
function or hypergeometric function.  Examining the integral for
large $\Omega$, we find the asymptotic space-time is the
flat vacuum in each theory.
Apart from the relation between
$\Omega$ and $y = \ln x$, Eq.(\ref{pscr}) is identical to that in
vacuum solutions in each theory.  Thus, the metric $g_{\alpha \beta}$
is flat in 2-dimensional sense and all the information regarding the
the curvature of space-time is contained in Eq.(\ref{psc}).
{}From Eq.(\ref{pscf}), we see that
$f$ diverges logarithmically near $\Omega = 0$ and, for large $\Omega$,
asymptotically falls off like $1/\Omega^{1+q}$.  This asymptotic
behavior, in $d$-dimensional Einstein gravity, translates to fall-off
like $1/r^{d-3}$ in terms of geometric radius $r$.  In between these
two limits, $f$ is a monotonically increasing function of $\Omega$.
The asymptotic solutions for large $\Omega$ can also be reproduced directly
by taking $G \rightarrow 0$ limit, where $G$ is the gravitational
constant.  This limit is the same as the weak field approximation
taking a vacuum metric in each theory as a fixed background metric.
The wave equation
for $f$ in $s$-wave sector under this fixed background geometry has
general static solutions of the form $ f = constant / \Omega^{1+q} +
f_1$, a solution describing a point scalar charge sitting at the
origin.  Thus, Eqs.(\ref{psc})-(\ref{pscr}) are the relativistic
generalization of the point scalar charge solution.  The strong
self-energy of the scalar field near the origin backreacts to the
space-time geometry and, in turn, this change softens the power-like
divergence of $f$ near the origin into milder logarithmic singularity.
We also note that the total $d$-dimensional
gravity-matter action in $d$-dimensional
Einstein gravity vanishes for
these solutions .  The physical region depicted in Fig.1c does not include
the point $\dy = 0$, a position where the horizon could have been formed.
This lack of trapped region implies the singularity at $\Omega = 0$
is naked. Obviously, black holes are not present in these solutions.

     In Fig.1d, we have generic cases of $M > 0$ and $f_0 >0$,
resulting $h_- > h_+ > 0$.  The field $f$ diverges logarithmically
near $\Omega = 0$ and the space-time at large $\Omega$ is asymptotically
flat.  Actually, $\F$ quite rapidly approaches the $\F$ for a black
hole with mass $M$ as $\dy$ gets large.
(See Appendix for the explicit form of $f$ in this case.)  The
space-time geometry is also rather similar
to the case of point scalar charges, Fig.1c; there are two values of
$\dy$, for which $\F$ vanishes.  The horizon $\dy = 0$ is
not included in the physical region to take the inverse of $\F$,
leading inevitably into naked singularities.  As a result, black
holes are not present in these solutions, although $M$ does not
vanish.  For a better physical understanding of this situation, let us
consider the case when $M >> f_0 > 0$, i.e., $h_- \simeq M/(1+q)$ and
$h_+$, a very small positive number proportional to $f_0$.  Then, we
might try to solve the wave equation for $f$ in fixed black hole
geometry with mass $M$, neglecting the backreaction of the self-energy
of $f$ on space-time geometry.  It is well known that the static
solutions for $f$ in this weak field approximation in curved space-time
diverge logarithmically at the black hole horizon, $\dy = 0$ in Fig.1d
(for example, $f \simeq \ln | r - 2M |$ in 4-d Einstein gravity
near the horizon $ r= 2M$).
Since the gravitational self-energy of $f$ correspondingly
diverges near the black hole horizon, the weak field approximation is not
valid there.  What our calculation shows instead is this large
self-energy near the horizon backreacts to space-time to cut off the black
hole from our view and produces a naked singularity in front of the
potential black hole horizon.
Indeed for small but finite $f_0$,
the width of the cusp near $\dy = h_+$ in Fig.1d gets very small but
finite.  Only for $f_0 = 0$, the cusp disappears and
we recover the black hole geometry.  In other words, when black holes
try to carry some scalar charge, they end up shutting themselves off
from our view, leaving a naked singularity.

     To summarize, other than some cosmological solutions with compact
range of the geometric radius, all solutions with asymptotically flat
space-time have naked singularities, except for black hole solutions
(and vacuum solutions) with identically vanishing scalar field $f$.
Thus, we completed the proof of no-hair theorem for all model theories
in our consideration.  Additionally we showed there are relativistic
analog of point scalar charges in each theory.

\newsection{Some Dynamic Solutions}

     The static point scalar charge solutions can, in some cases, be
simply extended to scattering solutions.  The $s$-wave sector of
4-d general relativity is such an example, as shown in this
section.  The solutions constructed in this fashion describe
the situations where either black holes are dynamically formed
as a result of the gravitational collapse or the incoming stress-energy
flux gets reflected toward the future (null) infinity.

\newsubsection{3.1. The Case of 4-d Einstein Gravity in $s$-wave Sector}

     The spherically symmetric reduction of 4-d Einstein gravity
coupled with a massless scalar field $f$
is the case when $\gamma = 2$, $\mu = -2$, $\lambda =1$ and
$\delta =1$\footnote{We note here that if we couple a massless fermion
field and describe the fermion $s$-wave sector via a bosonization
procedure, the only difference from the above specification of
parameters is the value of $\delta = 0$.} and, as our action shows, the
gravitational constant $G$ satisfies $16 \pi G = 1$.   Then,
the static solutions for $M = 0$, i.e. point scalar charge
solutions, are calculated to be
\begin{equation}
\Omega = \frac{e^{-2 \rho_0}}{4}  ( e^{4 \rho_0}(\ln (x/x_0 ) )^2 -
4 f_0^2 ) ,
\label{pointstatic}
\end{equation}
\[ \rho = \frac{1}{4} \ln \Omega - \frac{1}{2} \ln x  +\rho_0 , \]
\[
f = \ln ( \frac{ \sqrt{e^{2 \rho_0} \Omega + f_0^2} - f_0 }
{ \sqrt{e^{2 \rho_0} \Omega + f_0^2 } + f_0 } )  + f_1
= \ln ( \frac{ e^{2 \rho_0} \ln  (x/x_0 ) -2 f_0 }
{ e^{2 \rho_0 } \ln (x/x_0 ) + 2 f_0 } )
+ f_1  , \]
where $\rho_0$, $f_1$ and $x_0$ are arbitrary constants.
For simplicity, we take $x_0 =1$ and $\rho_0 = 0$
for further discussions.  Then,
as $\Omega \rightarrow \infty$, the behavior of $f$ asymptotically
approaches to $- 2f_0 /r + f_1 $
where the geometric radius $r$ is defined to be  $\sqrt{\Omega}$.
In this limit,
we find $r \rightarrow ( \ln x^+ +
\ln x^- ) / 2 $ and the 4-d metric becomes $ ds^2
\rightarrow  -
dx^+ dx^-/(x^+ x^- ) -
r^2 (d\theta^2 + \sin^2 \theta d\varphi^2 ) $.  After
a conformal transformation $\ln x^{\pm} \rightarrow x^{\pm}$, we
find that the asymptotic space-time is a flat Minkowskian.
We note that under weak field approximation taking a Minkowskian
metric as a fixed background metric, $s$-wave sector of static
solutions for $f$ is given by $ f = -2 f_0 / r + constant$.
The physical property of this solution
becomes more transparent if we use $r$ coordinate to describe
the geometry of longitudinal space-time instead of conformal
coordinates.  Then, from Eq.(\ref{pointstatic}), we find that
the metric is given by
\begin{equation}
ds^2 = dt^2 - \frac{1}{1+ f_0^2 / r^2} dr^2 - r^2
(d \theta^2 + \sin^2 \theta d \phi^2 ) .
\end{equation}
The above-mentioned asymptotic behavior is clear from this equation.

     The $M = 0$ case solutions are of particular importance
as we can generalize the static solutions to dynamic ones.
We note the structure of $s$-wave sector of 4-d Einstein
gravity is very simple as far as the limit $G \rightarrow 0$ is
concerned.  Then, we have a fixed Minkowski space-time and the linear
wave equation for $f$ has the general dynamic solution of the form
$ f = ( f_+ (x^+ ) + f_- (x^- ) ) /r $.  Comparing it to the static
solutions in the same limit, we find that replacing the constant $f_0$
in static solutions
with an arbitrary chiral field generates the general dynamic solutions.
This consideration suggests
that the scalar charge $f_0$, even in general relativistic cases,
may be a chiral field
instead of being a strict constant, similar to what happens in
$G \rightarrow 0$ limit.  In the framework of general relativity,
however, this simple extension is not possible
in general, since there can be a non-trivial corrections to
Eqs.(\ref{pointstatic}) of the order of $\partial_{\pm} f_0$.
In case of asymptotically steady incoming and outgoing stress-energy
flux, though, the corrections are easily found.
Forgetting about
global boundary conditions, the result of this extension is
\begin{equation}
\Omega = \frac{1}{4}
( \ln x )^2 - \frac{1}{4} ( k_+ \ln x^+ - k_- \ln x^-  + q_0 )^2 ,
\label{scattering}
\end{equation}
\[ \rho = \frac{1}{4} \ln \Omega - \frac{1}{2} \ln x
 + \frac{1}{2} \ln ( 1 +  k_+ k_- ) ,  \]
\[ f = \ln ( \frac{ \ln x -  (k_+ \ln x^+ - k_- \ln x^- + q_0 ) }
                  { \ln x +
                 (k_+ \ln x^+ - k_- \ln x^- + q_0 ) } ) + f_1 , \]
where $k_{\pm}$ and $q_0$ are constants.
We can straightforwardly verify that
these solutions satisfy field
equations derived from Eq.(\ref{caction}) and the corresponding
gauge constraints.  Apart from the additional constant term in the
expression
for $\rho$, the correction term originating from $\partial_{\pm} f$,
Eqs.(\ref{scattering}) are the same as
Eqs.(\ref{pointstatic}) with $f_0 = (q_0 + k_- \ln x^+
- k_+ \ln x^- ) /2 $.
The correction turns out to be rather simple in this case where
the charge $f_0$ has terms only up to linear terms in $\ln x^{\pm}$,
which are asymptotically flat conformal coordinates near the past or
future infinity.
The asymptotic stress-energy tensor averaged over the
transversal sphere
in a conformal coordinate system $v = \ln x^+$ ( $ u = \ln x^- $ )
that becomes asymptotically flat near the past (future) infinity
is calculated to be $T_{vv} =   k_{+}^2$ ($ T_{uu} = k_{-}^2$).  Thus,
$k_{\pm}^2$ are interpreted to be an incoming and an outgoing energy
flux, respectively.
The constant $q_0$ represents the background component of the scalar charge.
The $q_0 = 0$ case of Eqs.(\ref{scattering}) was reported in
mathematics literature \cite{dc} (named scale invariant solutions)
and in physics literature \cite{roberts} where the solutions were used
to investigate the violation of the cosmic censorship hypothesis.
For scattering situations where incoming and
outgoing flux can coexist, a slightly
generalized version (\ref{scattering}) proves
to be useful.  Additionally,
the presence of $q_0$ term enables us to consider the time evolution
of the (multiple) square-type incoming
energy pulses by successively gluing
our solutions, unless a black hole is formed in an intermediate stage.
To name a few other applications possible with our result, we
can construct various scattering solutions, cosmological solutions
and point particle solutions
with time-varying charge at the origin, depending on the boundary
conditions and initial conditions.

     To illustrate a black hole formation in this system, we
consider this physical situation; in the asymptotic (null) past,
we turn on the constant incoming stress-energy flux, inject it in a
spherically symmetric fashion for a time duration and turn it off.  After
a dynamical evolution, we will be interested in what comes out in the
asymptotic (null) future.  The physical region of space-time in our
consideration is specified by the requirement $\Omega \ge 0$, since
the angular coordinates should not have time-like signature.  Thus, the
natural boundary is the origin, $\Omega = 0$.  In the limit
$G \rightarrow 0$, with a
fixed Minkowskian background, the situation is represented by the
solutions depicted in Fig.2 and given by
\begin{eqnarray}
 &  f = 0                    & \mbox{I}    \label{fapp}  \\
 &  f = \frac{-kv}{r}         & \mbox{II}    \nonumber \\
 &  f = -2k                   & \mbox{III}   \nonumber \\
 &  f = \frac{-k(v_0 + u)}{r} & \mbox{IV}    \nonumber \\
 &  f = 0                    & \mbox{V}     \nonumber \\
 &  f = \frac{-kv_0 }{r}      & \mbox{VI}    \nonumber
\end{eqnarray}
where we use flat coordinates $v = \ln x^+$ and $u = \ln x^-$, thereby
$\Omega = r^2 = ((u+v)/2)^2$.

\begin{center}
\leavevmode
\epsfysize=7cm
\epsfbox{fig2.eps}
\end{center}
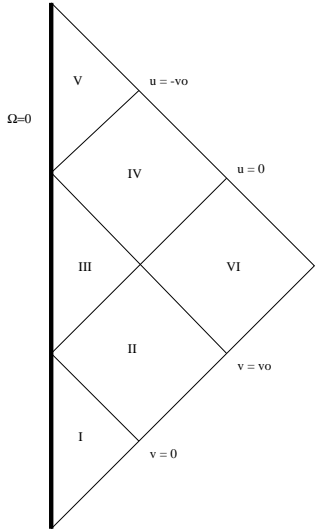
{\small Fig.2. The Penrose diagram shows the space-time region in fixed
Minkowskian background.  The bold line denotes the space-time boundary
$\Omega = 0$.  The constant incoming stress-energy flux is turned
on during the null time $v =0$ and $v = v_0$.}

The initial data on the past null
infinity were so chosen to describe the turn-on of the constant
incoming stress-energy flux $T_{vv} = k^2$ there at $v = 0$ and the
subsequent turn-off at $v = v_0 \ge 0 $.  We also imposed a boundary
condition by requiring the field $f$ to be finite along $\Omega =0$.
This approximate solution in a {\it fixed} Minkowskian
background has a property that any
unbounded amounts of the incoming stress-energy flux are
totally reflected off from the origin into the future
null infinity.  However,
this picture qualitatively changes as we consider
the exact solutions focusing on the space-time geometry change
due to the stress-energy of the scalar field.  Under the same choice
of the initial data on the past null infinity and the boundary
condition along $\Omega = 0$, we can construct the following
relativistic solutions.  For $\Omega$ field, we find
\begin{eqnarray}
 & \Omega =  \frac{1}{4}(u+v)^2               & \mbox{I}
\label{ssol} \\
 & \Omega =  \frac{1}{4}(u+v)^2 - \frac{1}{4} k^2 v^2     & \mbox{II}
           \nonumber \\
 & \Omega =  \frac{1}{4}(\sqrt{1 - k^2} v + \frac{u}{\sqrt{1-k^2}} )^2 &
           \mbox{III} \nonumber \\
 & \Omega =  \frac{1}{4}((1-k^2 ) v + \frac{u}{1-k^2} + k^2 v_0 )^2
            - \frac{1}{4} k^2 ( v_0 + \frac{u}{1-k^2} )^2 &
           \mbox{IV}  \nonumber \\
 & \Omega =  \frac{1}{4}((1-k^2 ) v + \frac{u}{1-k^2} + k^2 v_0 )^2 &
           \mbox{V} \nonumber
\end{eqnarray}
and
\begin{eqnarray}
 & f = 0    & \mbox{I} \label{fssol} \\
 & f = \ln ( \frac{ v + u - kv}{v+u + kv} ) & \mbox{II} \nonumber \\
 & f = \ln ( \frac{1-k}{1+k} )              & \mbox{III} \nonumber \\
 & f = \ln ( \frac{ (1-k^2 )v+u/(1-k^2 )+k^2 v_0 -k(v_0 +u/(1-k^2 )}
                { (1-k^2 )v+u/(1-k^2 )+k^2 v_0 +k(v_0 +u/(1-k^2 )} )
      & \mbox{IV} \nonumber \\
 & f = 0    & \mbox{V} \nonumber
\end{eqnarray}
for $f$ field.  The function $\rho$ can be read off from
Eqs.(\ref{scattering}).  We note the limit $v_0 \rightarrow \infty$,
in which we do not turn off the incoming flux, leaves us with the
regions I, II and III.  Our solutions in this limit were previously
discovered and discussed by some authors \cite{dc} \cite{brady}
\cite{kiem}.  The solution in the region VI
is rather complicated and, in generic cases, can not be obtained
using Eqs.(\ref{scattering})
or the conformally transformed version of them for the reasons
explained later.  However, we can obtain the region IV solution
by matching the III-IV boundary and from a reasonable physical
assumption that there is no infalling stress-energy flux in the region
IV since the asymptotic incoming stress-energy flux is turned-off
for $v > v_0$.  Eqs.(\ref{fssol}) reduce, after taking
the limit $G \rightarrow 0$ (or $k \ll 1$) and under the leading
order approximation, to Eqs.(\ref{fapp}) and the geometry
is almost Minkowskian, as expected.

\begin{center}
\leavevmode
\epsfysize=7cm
\epsfbox{fig3.eps}
\end{center}
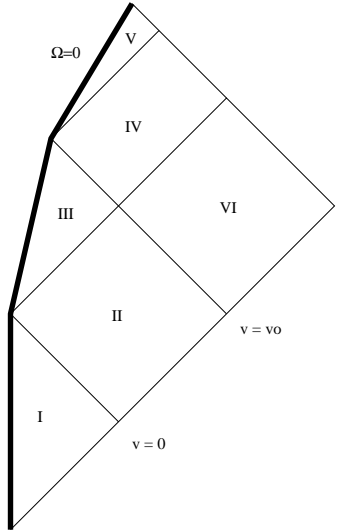
{\small Fig.3. The Penrose diagram for a typical subcritical
case.  The boundary $\Omega = 0$, the bold line shown above,
remains time-like throughout the entire space-time.}

     The region I, bounded by the past null infinity, $v=0$ and
the origin $u = -v$, represents the Minkowski space
before the turn-on of the constant incoming flux.
Regardless of $v_0$, the qualitative properties of our solutions
beyond the region I
are distinctively different for $k<1$ and $k>1$ (notice $T_{vv} = k^2$).

     For $k<1$, depicted in Fig.3, all the regions I-VI exist.
The region II, bounded by
$v = v_0$, $v = 0$ and the past null infinity, represents the
propagation of the incoming particles before any of them hits the
boundary $\Omega = 0$.  The region III represents the region where
incoming particles and outgoing particles reflected off from
$\Omega = 0$ coexist.  The net energy flux cancels as a result, and
we have a flat space-time throughout the region and $f$ field, a constant.
However, with respect
to the flat Minkowskian in the region I, the flat space-time
in the region III
is Lorentz boosted with relative speed $v/c = k^2 / ( 2 - k^2 )$.
The region IV contains the outgoing particles further
propagating toward the future null infinity.
In the asymptotic out region,
which includes the asymptotically flat future null infinity
in the region IV and the
strictly flat region V, there is additional
Lorentz boost with respect to the flat
region III with relative speed $v/c = k^2 / (2 - k^2 )$ and a
translation $k^2 v_0$.  Due to these
Lorentz boosts, the original energy pulse looks time-contracted
and amplified after the reflection off the origin from the point
of view of asymptotic in-observers.
On the other hand, from the point of view of
asymptotic out-observers who use $U(u) = u/(1-k^2 )$ and $V(v) =
(1- k^2 ) v$, the outgoing stress-energy flux is $T_{UU} = k^2$ and
the duration of the pulse is $v_0$, exactly the same as what
an asymptotic in-observer sent in.  Thus, the total outgoing energy
carried by the outgoing particles in the region IV is the same as the
total incoming energy thrown in initially.  From the conservation of
energy, we can infer that no further out-going energy flux comes out
through the future null infinity in the region VI.  We also note that
the amount of the translation of coordinates from the region I to
the regions IV and V, $k^2 v_0$, is proportional to the total injected
energy.  In summary, for $k <1 $, the path of the origin remains
time-like throughout the whole space-time and incoming particles
are scattered forward into the future null infinity, just as in the
subcritical regime of the numerical simulations.

\begin{center}
\leavevmode
\epsfysize=7cm
\epsfbox{fig4.eps}
\end{center}
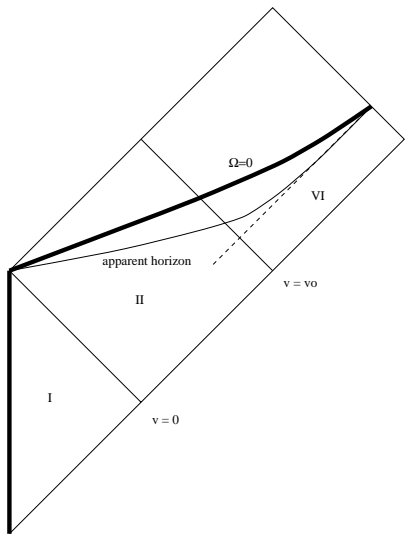
{\small Fig.4. The Penrose diagram for a typical supercritical
geometry.  The boundary $\Omega = 0$, the bold line, becomes
space-like for $v>0$ and the apparent horizon forms.  The details
shown in the region VI are conjectural.}

     If $k > 1$, as shown in Fig.4, the path of the origin becomes
space-like and form a trapped surface in the region II.  In this
case, the regions III, IV and V disappear and our solutions in the
region II become
exactly the same as the one obtained in \cite{dc}.
As explained in detail in \cite{dc}, the resulting space-time
is a dynamic black hole with increasing mass, for we
do not turn off the constant incoming flux up until $v= v_0$.  Thus, this
corresponds to the supercritical phase of this scattering system.
At $k =1$, which can thus be interpreted as a black hole formation
threshold, or a phase transition point, the path of the origin
becomes light-like in the region II.
As the numerical studies and the above considerations
suggest, the order parameter of this system is the black
hole mass, which vanishes for $k<1$ and becomes non-vanishing for
$k>1$.  Then, the important physical
quantity to compute, given our
exact solutions in supercritical regime, is the critical
exponent.
The geometric radius $r= r_A (v) $ of the apparent
horizon of the dynamic black hole in supercritical case,
determined by
$\partial_v r = 0$, is calculated to be
\begin{equation}
 r_A (v)= \frac{1}{2} (k- 1 )^{1/2} (k+ 1 )^{1/2} k v .
\label{mass}
\end{equation}
The $1/2$ times the value of this
corresponds to the apparent mass $M_A$
of the dynamic black hole.  The linear dependence on $v$ is
understandable as it is the time duration between the
turn-on of the incoming flux and the reference time $v$.
We also note that the angular integrated incoming energy
flux is $4\pi k^2= \frac{1}{4} k^2 /G$.
Defining the transition point $p^* = 1$ and $p = k$, we find
that the critical exponent in this case is $\Delta = 1/2$
in a scaling relation $M_A \simeq (p - p^* )^{\Delta}$.

     In the numerical study \cite{choptuik}, the scaling relation
is given for the asymptotically static black hole mass $M_{BH}$, measured
in the future (null) infinity, and $p$.  In our case, on the other hand,
since we do not know the solutions in the region VI, the scaling relation
between $M_A$ and $M_{BH}$ is not available.  However, it is possible
to show that, in generic cases, we can not choose static solutions in
the region VI (in the case of $G \rightarrow 0$ limit, the solutions
in the region VI are indeed static as shown in Eqs.(\ref{fapp})),
thereby identifying $M_{BH}$ and $M_A$.  To show this, we recall the
general static solutions
in the region VI are given by
\begin{equation}
U(u) + V(v) = 2 \int \frac{rdr}{F^{-1}_{h_- , h_+} (r )}
\end{equation}
from section 2 in conformal gauge.  Along II-VI boundary,
$v = v_0$, we have to satisfy
\begin{equation}
\Omega = \frac{1}{4} ( v_0 + u)^2 - \frac{1}{4} k^2 v_0^2 = (r(u))^2
\label{b1}
\end{equation}
and
\begin{equation}
\partial_v \Omega = \frac{1}{2} u  + \frac{1}{2} ( 1-k^2 ) v_0 .
\label{b2}
\end{equation}
By requiring $v$ coordinate to be $C^{\infty}$ along the past null
infinity, we find $V(v) = v$.  Eq.(\ref{b1}) determines the function
$U(u)$ via
\begin{equation}
U(u) + v_0  = 2 \int^{r = r(u)} \frac{rdr}{F^{-1}_{h_- , h_+ } (r)}
\end{equation}
where the function $r(u)$ is given in Eq.(\ref{b1}).  After some
calculations, Eq.(\ref{b2}) becomes
\begin{equation}
 \F (r) = \sqrt{ (r - \frac{1}{2} k^2 v_0 )^2 - \frac{1}{4} k^2
v_0^2 }
\label{conb}
\end{equation}
and this should determine $h_{\pm}$.  Unless the scalar charge
$kv_0$ goes to zero, Eq.(\ref{conb}) can not be satisfied for any
values of $h_{\pm}$.  Thus, we have shown the above statement
for generic cases.  The only exceptional cases are when the scalar charge
vanishes after $v = v_0$.  A notable example of such exceptions is the
shock wave
injections \cite{dray}.  In this case, we take $v_0 \rightarrow 0$ limit
for fixed $k^2 v_0$, to make the injected energy finite.  Then, we see that
the region II disappears and $kv_0$ vanishes as $v_0 \rightarrow 0$.
The exact solutions, obtained from the above consideration and shown in
Fig.5, are given by
\begin{equation}
\Omega = \frac{1}{4} (u + v)^2
\end{equation}
for the region I, a flat Minkowskian, and
\[  r + 2m \ln | \frac{r}{2m} -1 | = \frac{1}{2} (U(u) + v) \]
along with $U(u) = u + 4m \ln | u/4m -1 |$ and $m = k^2 v_0 /4$ in
the region VI, the Schwarzschild geometry.  The scalar field vanishes
in the whole region except
the line $v = 0$ (shock-wave).  Recalling Eq.(\ref{mass}) and
the divergence of $k$, this is an example of extremely supercritical
cases and we indeed have $M_A = M_{BH}$.

\begin{center}
\leavevmode
\epsfysize=7cm
\epsfbox{fig5.eps}
\end{center}
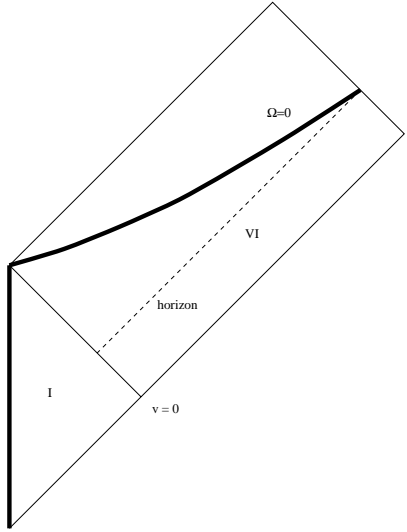
{\small Fig.5.  The Penrose diagram for a shock wave injection
along $v=0$.  The geometry for $v>v_0$ is described by Schwarzschild
metric and the geometry below that is flat Minkowskian.}

     In supercritical cases, we consequently infer that the
space-time geometry in the region VI goes through some
transient non-static period as long as the scalar charge does not vanish
for $v> v_0$.  The no-hair theorem supports this result.
For a given finite incoming energy, the theorem insists the only
possible candidates for the asymptotic final state geometry
are black holes with no scalar charge.  Thus the role of the transient
period is to bleach the static component of the residual scalar charge.
During this process, the apparent horizon that was
initially space-like at the turn-off time
will settle down to a future null direction, the asymptotic final
horizon, slightly changing its geometric radius.  The numerically
obtained $M_{BH} \simeq (p - p^* )^{\Delta}$ with $\Delta
\simeq 0.37$ is very difficult to calculate, as the scaling
relation between $M_{BH}$ and $M_A$ gets complicated through this
process.  In the subcritical cases, we showed that the energy conservation
prohibited the emission of out-going flux in the region VI.  In
contrast, the
numerically established scaling behavior for the asymptotic black hole
mass $M_{BH}$ implies that some energy flux escapes into the future
null infinity in the region VI for supercritical (especially near
critical) cases.

\newsubsection{3.2. The Case of Generic Gravity Theories}

     To apply our method of the extension to other gravity theories,
we have to get a dynamic solution in $G \rightarrow 0$ limit, taking a
vacuum metric as a fixed background.  For example, in $d$-dimensional
Einstein gravity, the linear wave equation in $(t,r)$ coordinates
in the limit is given by
\begin{equation}
 \partial_t^2 f^* - \partial_r^2 f^* + \frac{1}{2} (d-2)(d-4)
\frac{1}{r^2} f^* = 0
\end{equation}
where $f = f^* / r^{(d-2)/2} $ and $r$ is the geometric radius.
The vacuum solution for the
space-time satisfies $\Omega = r^{d-2}$.  We
immediately find that, unless $d=4$, the solutions are quite complicated.
Additionally, the static solutions in each case are $f = f_0 /
r^{d-3} + f_1$ where $f_0$ and $f_1$ are constants.  Thus, one can
imagine setting $f_0 = r^{(d-4)/2} f^*$ in our static solutions
for $f^*$ that produces the constant incoming (outgoing)
flux in the past (future) null infinity, for the relativistic extension.
Whether this procedure works
or not is not yet clear.

     In case of the CGHS model for pure gravity sector, or $d \rightarrow
\infty $ limit, the linear wave equation is
\begin{equation}
 \partial_t^2 f^* - \partial_x^2 f^* + \frac{k^2}{4} f^* = 0
\end{equation}
where $f = \exp ( -kx/2) f^*$.  Here the geometric background is the linear
dilaton vacuum $\Omega = \exp (k x)$ and $x$ is the asymptotically flat
spatial coordinate.  The major difficulty in this case is the
existence of mass term, causing $f^*$ to be non-chiral
even for the asymptotic infinities.

     One interesting point for theories with $d > 4$ (including
$d = \infty$ case) is that
$\sqrt{\Omega} f$ vanishes as $\Omega \rightarrow \infty$ for
non-trivial (non-constant) static solutions $f$.
This implies the static scalar charge component of
solutions completely decouples from the dynamic solutions; the
information about the static component can not be contained in the initial
data specified along the past null infinities, since it vanishes too fast.
The $d = \infty$ case shows this most clearly.  The physical frequency
spectrum of $f^*$ that is massive requires the wave length becomes
purely imaginary number if we set the frequency to be zero.  In fact,
the physically allowed frequency is larger than the
mass in this theory.  Thus,
the transient behavior of static scalar charge bleaching in 4-d Einstein
gravity after the turn-off of the incoming energy flux is expected to
be absent in theories with $d > 4$.

\newsection{Discussions}

     Our derivation of the general static solutions in section 2
relies heavily on the existence of three rigid symmetries of the
action.  Among these symmetries, there is a static remnant from the
underlying conformal invariance, which can thus be made local.  However,
the symmetry whose charge relates to a black hole mass does not share
this property.  In the CGHS model with $\delta = 0$, we have additional
rigid symmetry $\Omega \rightarrow  \Omega + A$ where $A$ is a constant.
Additionally, this symmetry and $f \rightarrow f + B$ can both be made
local.  These local symmetries resulting from the underlying 2-d Poincare
current algebraic symmetry are enough to determine the general solutions
of the CGHS model.  One natural question, then, for other gravity theories
is whether it is possible to find three rigid symmetries that can be
gauged as what happens in CGHS model.  The existence of them will be
helpful in finding the general solutions in each theory.

     The dynamic solutions in $s$-wave 4-d Einstein gravity obtained
in section 3 are interesting from many point of view.  First, the
dynamics of the point $\Omega = 0$ shows a remarkably similar behavior
to that of dynamical moving mirror considered in \cite{vc} for the CGHS
model.  Trajectories of both points dynamically reacts to the
incoming energy flux.  Furthermore,
depending on the asymptotic incoming energy flux, both
show the subcritical behavior where the trajectory is strictly time-like
and the supercritical behavior where the trajectory becomes space-like.
Thus, in some sense, the seemingly extraneous introduction of
reflecting dynamic boundary
(moving mirror) in the CGHS model makes the model behave more like
phenomenologically interesting $s$-wave sector of 4-d Einstein gravity.
In the latter theory, the dynamics of $\Omega = 0$ is a necessary
consequence of the theory, as we showed.  Second, it will be very
interesting to study the quantum field theory on a near critical but
subcritical geometry.  In this case, the incoming and outgoing null
coordinates are related by very large Lorentz boost, somewhat similar
to the infinite red shift near the black hole horizon.  Unlike the
black hole case, however, no information should be lost in the case
of subcritical solutions and, additionally, the mode expansion for
the free scalar field on the asymptotic (null) infinities are well
defined.  Therefore, the near critical and subcritical solutions
provide a nice background geometry to study gravitational interactions
between incoming and outgoing quantum fluctuations.

     Our third interest lies in the phase transition itself.
The black hole phase transition in the CGHS model is not as
difficult as in the spherically symmetric 4-d Einstein gravity.
The complicated transient
behavior is absent in this case, due to the simplified
dynamics of the model.  Therefore, it
would be possible to directly glue a static
dilatonic black hole at $v=v_0$ and thereby getting a relation
$M_A \simeq M_{BH}$ if we had considered the CGHS model from the
the outset.  An interesting observation in this regard is
the critical exponent 0.5
for the scaling relation for the apparent mass
in $s$-wave 4-d Einstein gravity is the
same as the critical exponent obtained
by Strominger and Thorlacius in the CGHS model \cite{strominger}.
There is a reason for this connection as suggested in
\cite{birkhoff}.  The pure gravity sector of the CGHS model,
other than being a target
space effective action from string theory,  can be considered
as a leading order theory in the $1/d$-expansion of the
spherically symmetric $d$-dimensional Einstein gravity.
We can, therefore, adopt $1/d$-expansion  and consider the leading
order behavior in the description of the
complex transient process in 4-d Einstein gravity.
As a zeroth order approximation, we glue a CGHS black hole
directly to our solutions in the region II to deduce
the approximate scaling relation $M_A  \simeq M_{BH}$. (See section
3.2.)  Thus, the leading
order approximation of the exact critical exponent for square
pulse-type incoming energy flux is
now calculated to be $0.5$.
Since the next order correction to the critical exponent is
expected to be an order of $1/d = 0.25$ and the numerically
calculated value is about 0.37, our leading order value, $0.5$,
seems plausible.
Furthermore, it is conceivable
that the exact critical exponent ($M_{BH} \simeq |p - p^* |^{\Delta}$)
for pulse-type incoming
energy flux  in
$d$-dimensional spherically symmetric Einstein gravity for $d >4$
would be  0.5, considering the presumed lack of transient
behavior discussed in section 3.2.
In many other
cases of phase transitions in condensed matter physics, the critical
exponent gets the scaling violation for only lower dimensional
cases.  It will be an interesting
exercise to verify this conjecture and, additionally,
develop a systematic
perturbation theory with a dimensionless
expansion parameter $1/d$
to tackle other difficult problems in 4-dimensional gravity.

\newsubsection{Acknowledgement}

     The author wishes to thank H. Verlinde and D. Christodoulou
for useful discussions and comments.

\appendix
\newsubsection{Appendix. The Alternative Derivation of Static Solutions}
     Instead of choosing a conformal gauge, we can choose a
gauge used in \cite{kiem} where the general solutions of pure
gravity sector in our consideration were obtained to prove
Birkhoff's theorem.  From now on, we follow the convention
of the reference.  The metric tensor in this gauge is given
by
\begin{equation}
g_{\alpha \beta} =  \left[ \begin{array}{cc}
           - \alpha^2    &   0       \\
             0           & \beta^2   \end{array} \right]   .
\end{equation}
Furthermore, we choose coordinates in such a way that
$x^1 = \Omega = \exp ( -2 \phi ) $ and require $[ \partial_0
, \partial_1 ] = 0$.  We note that the signature convention,
the time component being negative, is different from section
2 to closely follow the reference \cite{kiem}.

     The resulting equations of motion can be written as
\begin{equation}
\frac{\partial_1 \alpha}{\alpha} + \frac{\gamma}{8 \Omega}
- \frac{\beta^2}{2} \Omega V( \Omega )
- \frac{\Omega^{\delta}}{4} [ ( \partial_1 f )^2
+ \frac{\beta^2 }{\alpha^2} ( \partial_0 f )^2 ] = 0
\label{a1}
\end{equation}
\begin{equation}
\frac{\partial_1 \beta}{\beta} + \frac{\gamma}{8 \Omega }
+ \frac{\beta^2}{2} \Omega V( \Omega )
- \frac{\Omega^{\delta}}{4} [ ( \partial_1 f )^2
+ \frac{\beta^2 }{\alpha^2} ( \partial_0 f )^2 ] =0
\label{a2}
\end{equation}
\begin{equation}
 \partial_1 ( \frac{\alpha}{\beta} \Omega^{\delta} \partial_1
f ) - \partial_0 ( \frac{\beta}{\alpha} \Omega^{\delta}
\partial_0 f ) = 0
\label{af}
\end{equation}
along with a gauge constraint
\begin{equation}
\frac{\partial_0 \beta}{\beta } - \frac{1}{2} \Omega^{\delta}
\partial_0 f \partial_1 f = 0 .
\label{agauge}
\end{equation}
Since we are interested in getting general {\it static}
solutions, we require $\partial_0 f = 0$ and $\partial_0 \beta
= 0$.  Thus the gauge constraint (\ref{agauge}) is automatically
satisfied.  Furthermore, Eq.(\ref{af}) can be directly integrated
to yield
\begin{equation}
 \Omega^{\delta} \partial_1 f = f_0 \frac{\beta}{\alpha} .
\label{aintf}
\end{equation}
Adding (\ref{a1}) and (\ref{a2}) gives
\begin{equation}
\partial_1 \ln A + \frac{\gamma}{4 \Omega}
- \frac{f_0^2}{2 \Omega } \frac{1}{B^2} = 0
\end{equation}
while subtracting them produces
\begin{equation}
\partial_1 \ln B = \mu \Omega^{1 - \lambda} \frac{A}{B}
\end{equation}
where we define $A = \alpha \beta $ and $B = \alpha / \beta$.
We also set $\delta = 1$ and $V( \Omega ) = \mu \Omega^{-
\lambda}$.  Introducing $\bar{A}$ via $A = \Omega^{\lambda -2}
\bar{A}$, we can rewrite above equations as follows.
\begin{equation}
\frac{d}{d \phi} B = -2 \mu \bar{A}
\label{a3}
\end{equation}
\begin{equation}
\frac{1}{\bar{A}} \frac{d}{d \phi} \bar{A} = - ( 2 + 2q
+ \frac{f_0^2 }{B^2} )
\label{a4}
\end{equation}
Here, as in section 2, $q$ is defined as $q = 1 - \lambda - \gamma
/4$.  We can plug Eq.(\ref{a3}) into Eq.(\ref{a4}) and
straightforwardly integrate it once to obtain
\begin{equation}
- \frac{d}{d\phi} B = \frac{1}{B} ( 2 (1+q ) B^2 + 2M B - f_0^2 )
\end{equation}
where $M$ is the constant of integration.  After integrating
this once more, we finally get
\begin{equation}
k \Omega^{1+q} = | h_+ - B |^{\frac{h_+}{h_+ + h_-}}
                 | h_- + B |^{\frac{h_-}{h_+ + h_-}}
               = F_{h_- , h_+ } ( B )
\label{ab}
\end{equation}
where $k$ is a constant of integration and
\[ h_{\pm} = \frac{1}{2(1+q) } ( \sqrt{M^2 + 2(1+q) f_0^2 }
\mp M ) . \]
We note that Eq.(\ref{ab}) is exactly the same as Eq.(\ref{omega1})
from section 2 if $B$ is replaced by $\frac{d \Omega}{dy}$.
Using Eq.(\ref{a3}), $A$ can be determined as
\begin{equation}
A = \frac{\Omega^{\lambda -1 }}{\mu} \frac{dB}{d\Omega} .
\label{aa}
\end{equation}
Eqs. (\ref{ab}) and (\ref{aa}) are our main results.  They implicitly
solve the metric in terms of $\Omega$.  Then $f$ can be directly
integrated using Eq.(\ref{aintf}).

     Choosing $B$ as a spatial coordinate in favor of $\Omega$ gives
more explicit solutions.  Then, the metric is computed to yield
\begin{equation}
 ds^2 = - \frac{B\Omega^{\lambda -1} }{\mu} ( \frac{d\Omega}{dB} )^{-1}
   dt^2 +  \frac{\Omega^{\lambda -1}}{\mu B} (\frac{d\Omega}{dB} ) dB^2
\end{equation}
where the function $\Omega$ in terms of $B$ is given in Eq.(\ref{ab}).
Using Eqs.(\ref{aintf}) and (\ref{ab}), the scalar field $f$ can be
explicitly solved as
\begin{equation}
f = \frac{f_0}{\sqrt{M^2 + 2(1+q) f_0^2}} \ln
  | \frac{B -h_+}{B+ h_- } | + f_1 .
\end{equation}
Setting $B = \frac{d \Omega}{dy}$ and $M=0$ in the above equation
reproduces Eq.(\ref{pscf}) in section 2.  The solutions of \cite{janis}
are recovered if we set $1+q = 1/2$ and $\lambda = 1$, the case of
4-d Einstein gravity, as long as $M \ne 0$.  We see that it is again
possible to derive the {\em general} static solutions.

\end{document}